\documentclass[twocolumn,tighten]{aastex62}
\usepackage{amssymb, amsmath,framed}

\usepackage{bm}
\expandafter\ifx\csname package@font\endcsname\relax\else
 \expandafter\expandafter
 \expandafter\usepackage
 \expandafter\expandafter
 \expandafter{\csname package@font\endcsname}
\fi
\def\bq{\begin{equation}}
\def\eq{\end{equation}}
\def\bqy{\begin{eqnarray}}
\def\eqy{\end{eqnarray}}
\begin{document}
\title{\large{Implications of Captured Interstellar Objects for Panspermia and Extraterrestrial Life}}

\correspondingauthor{Manasvi Lingam}
\email{manasvi.lingam@cfa.harvard.edu}

\author{Manasvi Lingam}
\affiliation{Institute for Theory and Computation, Harvard University, Cambridge MA 02138, USA}
\affiliation{Harvard-Smithsonian Center for Astrophysics, Cambridge, MA 02138, USA}

\author{Abraham Loeb}
\affiliation{Institute for Theory and Computation, Harvard University, Cambridge MA 02138, USA}
\affiliation{Harvard-Smithsonian Center for Astrophysics, Cambridge, MA 02138, USA}

\begin{abstract}
We estimate the capture rate of interstellar objects by means of three-body gravitational interactions. We apply this model to the Sun-Jupiter system and the Alpha Centauri A\&B binary system, and find that the radius of the largest captured object is a few tens of km and Earth-sized respectively. We explore the implications of our model for the transfer of life by means of rocky material. The interstellar comets captured by the ``fishing net'' of the Solar system can be potentially distinguished by their differing orbital trajectories and ratios of oxygen isotopes through high-resolution spectroscopy of water vapor in their tails.\\
\end{abstract}

\section{Introduction}
A few months ago, the first ever interstellar object, namely the comet `Oumuamua, was detected by the Pan-STARRS telescope \citep{MWM17,MF18}. The detection of `Oumuamua has led to several follow-up studies that aim to understand its origin, structure and travel time \citep{Mam17,GWK17,YZ17,JLR17,BS17,BWF17,FS18}. There have also been several studies devoted to understanding the implications for planet formation and architecture \citep{TR17,LB17,RA17,JT17,HZ17,Cuk18,Raf18}. The total number of interstellar objects comparable in size to `Oumuamua has been estimated to be $\sim 10^{15}\,\mathrm{pc}^{-3}$ \citep{DTT18}, which is much higher than some previous estimates \citep{Fra05,MTL09,EJ17}. As a result, it is worth exploring the implications of the updated number density for the capture of interstellar objects and the prospects for lithopanspermia, the possibility that life may have been transferred to Earth (and other habitable planets) by means of rocky material.\footnote{In addition to interstellar panspermia by means of ejected comets and asteroids, we observe that life could have been transferred via grains propelled by collisions with space dust \citep{Ber17} or radiation pressure \citep{Nap04}.}

In this paper, we shall explore some of the implications of the detection of `Oumuamua for the capture of interstellar asteroid and comets by the Solar system, and by binary stellar systems such as $\alpha$-Centauri. Subsequently, we will discuss the implications of our results in the context of lithopanspermia, the origin of life and macroevolutionary processes. We will also briefly explore the ramifications in searching for extraterrestrial life within the Solar system.

\section{Is it Easier to Detect Smaller or Larger Interstellar Objects?}
We denote the radius of the interstellar object by $R$, and the average number density of objects of radius greater than $R$ by $n(>R)$.\footnote{Gravitational focusing by the Sun introduces a small correction to the mean density of interstellar objects, given their velocity dispersion of tens of km/s at a distance of $\gtrsim 1$ pc from the Sun.} If we assume a power-law distribution with an exponent $\alpha$, the expression for $n$ is:
\begin{equation} \label{numbden}
    n(>R) \sim 2 \times 10^{15}\,\mathrm{pc}^{-3}\,\left(\frac{R}{100\,\mathrm{m}}\right)^{-\alpha},
\end{equation}
and we have employed the latest estimates based on the discovery of `Oumuamua \citep{Gai17,DTT18}. The spectral index $\alpha > 0$ remains unknown, and it may be necessary to employ a broken power-law to encompass objects much larger than `Oumuamua. Although the exact position of an object from the Earth will depend on its trajectory and the specific time at which it is detected, the mean distance $d$ (from the Earth) can be computed via
\begin{equation}\label{drel}
    d \sim \left(\frac{4\pi}{3}n\right)^{-1/3} \sim 1\,\mathrm{AU}\,\left(\frac{R}{100\,\mathrm{m}}\right)^{\alpha/3}.
\end{equation}

There are two central avenues through which radiation from the interstellar object can be detected. The flux density $S_\mathrm{max}$ at the blackbody peak is defined as
\begin{equation} \label{FluxD}
    S_\mathrm{max} \approx \frac{P}{4\pi d^2 \nu_\mathrm{max}},
\end{equation}
where $P$ is the emitted power and $\nu_\mathrm{max} \propto T_s$ with $T_s$ denoting the surface temperature of the object. The first avenue that we consider involves detecting reflected sunlight. In this scenario, we observe that
\begin{equation} \label{SolPow}
    P \propto L_\odot \left(\frac{R}{d + d_0}\right)^2,
\end{equation}
\begin{equation}
    T_s \propto L_\odot^{1/4} \left(d+d_0\right)^{-1/2},
\end{equation}
where we have introduced the notation $d_0 = 1$ AU and assumed that the albedo is independent of $R$. Substituting the above two relations into (\ref{FluxD}), we find 
\begin{eqnarray}
     S_\mathrm{max} &\propto& R^{(12-7\alpha)/6} \quad \mathrm{for} \quad d \gtrsim 1 \mathrm{AU} \nonumber \\
      S_\mathrm{max} &\propto& R^{(6-2\alpha)/6} \quad\,\, \mathrm{for} \quad d \lesssim 1 \mathrm{AU}.
\end{eqnarray}
Hence, for objects with $R > 100$ m, the condition $\alpha > 12/7$ would imply that the detection of smaller objects is easier since they will be more numerous and are therefore more likely to be found at closer distances. This criterion is likely to be satisfied for interstellar objects since the typical values of $\alpha$ for asteroids, Kuiper Belt Objects (KBOs) and comets range between $2$ to $5$ \citep{MTL09}. In contrast, if we consider objects with $R < 100$ m, the cutoff value becomes $\alpha = 3$. Based on the evidence from the Solar system, the value of $\alpha$ may lie either below or above this threshold, and hence it is not possible to arrive at unambiguous conclusions in this regime.

Next, we can consider the case where the emitted radiation is a consequence of radiogenic heating. In this case, we obtain the relations \citep{LL17}:
\begin{equation}
    P \propto R^\gamma, \quad T_s \propto R^{(\gamma-2)/4}
\end{equation}
where $\gamma \approx 3.3$ when the mass of the object is less than the Earth \citep{SGM07}. Substituting these scalings into (\ref{FluxD}), we arrive at
\begin{equation}
    S_\mathrm{max} \propto R^{(9-2\alpha)/3}.
\end{equation}
Hence, for cases with $\alpha < 9/2$ (which is mostly valid in our Solar system), larger objects would be easier to detect through this avenue even though they are fewer in number, and typically located at much farther distances.  

At this stage, it should be noted that the energy flux from the Sun dominates over the radiogenic flux unless the object is very far from the Sun, and is also sufficiently large. Thus, it may be easier to detect smaller objects illuminated by sunlight, whereas free-floating planets and/or objects very far from the Sun could be detected through their internal heating \citep{LL17}. Note that the difference in the two modes of detection is that the former case exhibits a $d^{-4}$ falloff whereas the latter has a $d^{-2}$ decline; the situation is analogous to objects that possess natural or artificial illumination respectively \citep{LT12}. The transition between these two regimes can be approximately determined by equating their respective values of $S_\mathrm{max}$, and we obtain
\begin{equation}
    R \sim 7.6\,\mathrm{km}, \quad d \sim 76\,\mathrm{AU},
\end{equation}
where we have assumed an albedo of $\approx 0.1$ and $\alpha \approx 3$.

We next consider whether an interstellar object can be resolved by current telescopes. We find that
\begin{equation}
    \theta = \frac{R}{d} \approx 0.14\,\mathrm{mas}\,\left(\frac{R}{100\,\mathrm{m}}\right)^{1-\alpha/3},
\end{equation}
upon using (\ref{drel}), implying that the object cannot be resolved except by reflected radar with the largest radio telescopes. For comets within our Solar system, which typically have $\alpha \approx 3$, the dependence of $\theta$ on $R$ will be very weak. We reiterate that our discussion of $S_\mathrm{max}$ and $\theta$ operate under the assumption that $d$ and $R$ are linked through (\ref{drel}), but the former depends on the location of the object at the time of observation. 

We also refer the reader to \citet{DTT18} and \citet{RA17} for further analyses concerning the detectability of interstellar objects that incorporate the effects of proper motion and the implications for planet formation, respectively.

\section{Capture and collisions with interstellar objects}\label{SecCap}
Next, we will present simple estimates for the capture rate of interstellar objects that can be captured by stellar systems, and also estimate the number of collisions with the Earth over its entire history.

\subsection{Capture of interstellar objects by stellar systems}\label{SSecCap}
We consider the Solar system and the $\alpha$-Centauri binary system as separate cases.

\subsubsection{Solar system}
In a seminal paper, \citet{Heg75} presented a comprehensive analytical treatment for the dynamical interaction of binary systems with field objects. Subsequently, this analysis was extended by several authors to estimate the rate of capture of interstellar comets \citep{Val83,Tor86,PD93,CRG16}. The capture rate of comets per year $\dot{N}$ is approximately given by
\begin{equation} \label{Caprate}
    \dot{N} \approx n \int_0^\infty V\,\sigma(V) f(V)\,dV,
\end{equation}
where $V$ is the velocity of the interstellar object, $f(V)$ represents the speed distribution of the objects and $\sigma(V)$ is the velocity-dependent capture cross section \citep{Has76}. The final result for a Maxwellian distribution can be expressed as follows:
\begin{eqnarray} \label{CapRate}
    \dot{N}\, &\approx&\,\, 2 \times 10^{4}\,\mathrm{yr}^{-1}\,\left(\frac{\epsilon}{0.1}\right) \left(\frac{n}{10^{15}\,\mathrm{pc}^{-3}}\right) \left(\frac{a}{10\,\mathrm{AU}}\right)^{1/2} \\
    && \times \left(\frac{v_\infty}{20\,\mathrm{km/s}}\right)^{-3} \left(\frac{m_1}{M_\odot}\right)^2 \left(\frac{m_2}{M_\odot}\right)^2 \left(\frac{m_1 + m_2}{M_\odot}\right)^{-5/2}, \nonumber
\end{eqnarray}
where $m_1$ and $m_2$ are the masses of the two binary objects, $a$ represents their separation, and $v_\infty$ is the characteristic velocity of the interstellar object \citep{Val83}. The efficiency factor $\epsilon$, which is typically $\sim 0.1$, has been introduced since the analytical result obtained by using the formulation developed by \citet{Heg75} is about an order of magnitude higher than the results from numerical simulations \citep{VI82}. Many of the captured objects will be characterized by very elliptical orbits and have a high likelihood of being ejected subsequently as discussed later.

Upon applying (\ref{CapRate}) to the Solar system with Jupiter and the Sun representing the binary, we arrive at
\begin{equation} \label{NdotJ}
    \dot{N}_\odot \approx 1.3 \times 10^{-2}\,\mathrm{yr}^{-1}\,\left(\frac{n}{10^{15}\,\mathrm{pc}^{-3}}\right),
\end{equation}
which is reasonably consistent with the numerical results discussed in \citet{PD93}. Another method for estimating $\dot{N}$ for the Jupiter-Sun system is via the formula
\begin{equation}
    \dot{N} \sim n \langle{\sigma}\rangle v_\infty,
\end{equation}
where $\langle{\sigma}\rangle$ represents the velocity-averaged capture cross-section of Jupiter. Using the data from Table 1 of \citet{Mel03} for $\langle{\sigma}\rangle$, we obtain $\dot{N}_\odot \approx 1.2 \times 10^{-2}\,\mathrm{yr}^{-1}$ for $n \sim 10^{15}\,\mathrm{pc}^{-3}$ and $v_\infty \sim 20$ km/s; this result is in excellent agreement with Eq. (\ref{NdotJ}).\footnote{On the other hand, using the semianalytic cross-section proposed in \citet{Tor86} leads to an underestimated value of $\dot{N}_\odot \approx 1.6 \times 10^{-6}\,\mathrm{yr}^{-1}$.} Thus, over the history of the Solar system, a total of $N \sim 6 \times 10^7$ objects could have been captured based on Eq. (\ref{NdotJ}). However, not all of the captured objects will remain in bound orbits. Numerical simulations suggest that comets are ejected over mean timescales of $t_{ej} \sim 4.5 \times 10^5$ yrs \citep{LD94}. Thus, only a fraction $\sim 10^{-4}$ of the interstellar objects are expected to be present in the Solar system at any given time, bringing the number down to $\sim 6 \times 10^3$.

From Eqs. (\ref{numbden}) and (\ref{NdotJ}), we can determine the maximum radius of the interstellar comet or asteroid $R_\mathrm{max}$ such that at least one such object will be (temporarily or permanently) captured during the history of the Solar system; this amounts to the condition $\dot{N}_\odot \tau \sim 1$, where $\tau$ is the age of the Solar system. Thus, we obtain
\begin{equation}
    R^\mathrm{max}_\odot \sim 0.1 \times \left(1.2 \times 10^8\right)^{1/\alpha}\,\mathrm{km}.
\end{equation}
As an example, if we consider elliptical comets with $\alpha \approx 2.9$, we find that $R^\mathrm{max}_\odot \approx 60$ km, which is about half the radius of Enceladus.  In contrast, if we use $\dot{N}_\odot t_{ej} \sim 1$ and solve for $R_\mathrm{max}$, we obtain a much smaller value of $R^\mathrm{max}_\odot \approx 2.6$ km. 

In contrast, if we consider the capture of approximately Earth-sized objects, it has been proposed that $n\left(> R_\oplus\right) \sim 10-100\,\mathrm{pc}^{-3}$ is possible \citep{SBMB,BQ17}. Upon substituting this estimate into (\ref{NdotJ}) and using the fact that the total lifetime of the Sun is $t_\odot \sim 10^{10}$ yr, we find that $\dot{N} \tau_\odot \sim 10^{-3}-10^{-4}$. Hence, this would imply that the fraction of solar-type stars that would capture an Earth-sized object within their lifetime is $\sim 10^{-3}-10^{-4}$. This result is in excellent agreement with the value of $\sim 10^{-4}$ obtained from numerical simulations, as seen from Sec. 6 and Fig. 10 of \citet{GP18}.

We will briefly delineate some of the potential methods that may be used for discerning passing interstellar objects from those formed within our Solar system.
\begin{itemize}
    \item Hyperbolic orbits could be partly indicative of interstellar origin, but they are not a good diagnostic by themselves since there are other mechanisms which can give rise to this phenomenon as well \citep{KD17}. However, their rapid motion across the sky ($200$ arcsec/hr) could serve to distinguish them from Solar system objects \citep{CRG16}.
    \item Objects approaching the Solar system can be characterized by their radiants and barycentric velocities. It has been proposed that interstellar objects display distinctive signatures with regards to the latter, i.e. the magnitudes of their inbound velocities are anomalously high \citep{DDA18}. This method has been used to identify eight candidates, such as comet C/2017 K2 \citep{DD18}, that may be of interstellar origin.
\end{itemize}
We discuss potential methods of detecting captured interstellar objects within our Solar system in Sec. \ref{SSecFish}.

\subsubsection{Alpha Centauri A \& B}
We can also carry out similar calculations using the $\alpha$-Centauri binary system. Upon using (\ref{CapRate}) with the appropriate parameters, we find
\begin{equation}\label{CapAB}
    \dot{N}_\mathrm{Cen} \sim 5.4 \times 10^3\,\mathrm{yr}^{-1}\,\left(\frac{n}{10^{15}\,\mathrm{pc}^{-3}}\right),
\end{equation}
which is approximately five orders of magnitude higher than (\ref{NdotJ}). Using the condition $\dot{N}_\mathrm{Cen} \tau_\mathrm{Cen} \sim 1$ (but ignoring possible ejection), where $\tau_\mathrm{Cen}$ is the age of the $\alpha$-Cen system, we can compute the corresponding value of $R^\mathrm{max}$, and it yields
\begin{equation}\label{RCen1}
    R^\mathrm{max}_\mathrm{Cen} \sim 0.1 \times \left(5 \times 10^{13}\right)^{1/\alpha}\,\mathrm{km}.
\end{equation}
If we use $\alpha \approx 2.9$ as before, we find $R^\mathrm{max}_\mathrm{Cen} \sim 0.8 R_\oplus$, thus indicating that even Earth-sized planets could potentially be captured. However, post-capture, the stability of the ensuing orbits is not an easy question to resolve although several analyses have identified regions for stable orbits \citep{WH97,AM14,QLK}. The reason stems from the fact that the stability depends on the eccentricity, inclination and nature (S-type and P-type).

\citet{QL16} conducted a detailed stability analysis of different types of orbits, and studied the fraction of test particles that survived over Gyr timescales. From Fig. 6(a) of this work, it can be seen that the fraction of test particles on circumstellar orbits that survive over Gyr timescales is subject to variations depending on whether the orbits are prograde or retrograde as well as their eccentricity. However, in each of these cases, the fraction $f_c$ of remaining curves flattens and approaches the value $f_c\sim 0.3$ for Gyr timescales. Thus, upon using (\ref{CapAB}) and a timescale of $1$ Gyr, we arrive at a total of $1.1 \times 10^{13}$ objects. Of these, a total of $N_0 \sim 3.2 \times 10^{12}$ objects are expected to survive on Gyr timescales after making use of $f_c \sim 0.3$. 

Note that $N_0$ refers to the number of captured objects with $R \gtrsim 100$ m. The maximum radius $R^\mathrm{max}_\mathrm{Cen}$ among these objects that survives to Gyr timescales is
\begin{equation}\label{RCen2}
 R^\mathrm{max}_\mathrm{Cen} \sim  0.1 \times \left(3.2 \times 10^{12}\right)^{1/\alpha}\,\mathrm{km},
\end{equation}
and using $\alpha \approx 2.9$ leads us to $R^\mathrm{max}_\mathrm{Cen} \sim 0.3 R_\oplus$; this radius is approximately midway between that of the Moon and Callisto. The difference between (\ref{RCen1}) and (\ref{RCen2}) is that the former represents the largest temporarily captured object whereas the latter is the largest captured object that survives over Gyr timescales. Hence, in the $\alpha$-Centauri binary system, Moon-sized interstellar objects that were captured may currently exist in certain circumstellar orbits. From Fig. 7 of \citet{QL16}, it can be seen that the apastron distance for most of the surviving test particles is less than $3$ AU and $5$ AU in the prograde and retrograde cases respectively. 

The transit depth is estimated via $\delta \approx \left(R_p/R_\star\right)^2$, where $R_p$ and $R_\star$ are the radius of the object and the star respectively. Hence, choosing $R_p \sim 0.3\,R_\oplus$ and $R_\star \approx R_\odot$ leads us to conclude that the transit depth for the largest object will be $\sim 10\%$ of the corresponding value for the Earth as it transits the Sun. The fact that an object of mass $\sim 0.02 M_\oplus$ has not been detected around either of the binary stars to date is along expected lines.\footnote{Note that the mass has been computed using the mass-radius relation $M_p \propto R_p^{3.3}$ for rocky or icy objects that are smaller than the Earth \citep{SGM07}.} This is because the current detection thresholds can only rule out worlds in the habitable zone with masses greater than $53 M_\oplus$ and $8 M_\oplus$ for $\alpha$-Centauri A and B respectively \citep{ZF18}.

The capture of Earth-sized interstellar objects by binary systems akin to $\alpha$-Centauri A and B may offer an alternative to conventional planet formation through accretion since the latter scenario is expected to be unfavourable as per some theoretical models \citep{TMS08}; see, however, \citet{XZG}.

\subsection{Collisions of interstellar objects and Earth}
The Sun is known to intercept comets at a rate given by Eq (\ref{Caprate}), and $\sigma$ represents the cross-sectional area of the interaction region. \citet{ZV99} undertook numerical simulations by taking into account a suitable speed distribution function and cross-section. Using the parameters from Sec. 10 of \citet{ZV99}, we obtain
\begin{equation} \label{CollRate}
    \dot{N}_c \sim 2 \times 10^{-5}\,\mathrm{yr}^{-1}\,\left(\frac{n}{10^{15}\,\mathrm{pc}^{-3}}\right),
\end{equation}
where $\dot{N}_c$ is the rate of collisions with the Earth. If we impose the condition $\dot{N}_c\, \tau \sim 1$, we can determine the maximum radius of the interstellar object that could have collided with the Earth during its history:
\begin{equation} \label{MaxImp}
    R^\mathrm{max}_c \sim 0.1 \times \left(9 \times 10^4\right)^{1/\alpha}\,\mathrm{km},
\end{equation}
and using $\alpha \approx 2.9$ yields $R^\mathrm{max}_c \sim 5$ km. Upon using (\ref{numbden}) and (\ref{CollRate}), we find that a kilometer-sized object may strike the Earth over a typical timescale of $t_0 \sim 10-100$ Myr; note that $t_0$ is dependent on $\alpha$.

\section{Potential biological implications}
We shall briefly discuss some of the possible roles that interstellar objects can play with regards to the origin and evolution of life on habitable planets. Given that the Solar system has a large pool of ``native'' asteroids and comets, interstellar objects may be expected to play a sub-dominant role. However, what should be noted is that the inventory of native objects may vary widely from one planetary system to another. For example, it is known that the amount of material orbiting the G8V star $\tau$ Ceti is more than that of the Sun by an order of magnitude \citep{GW04}. Thus, our discussion in this Section is likely to be more pertinent for planetary systems that have much less native material with respect to the Solar system \citep{ML13}.

\subsection{Impact on macroevolutionary processes}
Since the 1980s, several paleontological studies have identified approximately periodic patterns in the fossil diversity record \citep{HW97,Bam06}, with two of the most widely analyzed timescales being $26$ Myr \citep{Ra84} and $62$ Myr \citep{RM05}. It is evident that these two timescales are on the same order as $t_0$, i.e. the timescale for the impact of kilometer-sized interstellar objects. We observe that this apparent coincidence of timescales shares some similarities with previous theories regarding cometary impacts \citep{BJ09}, but the latter class of models either relied on the Sun's passage through the Milky Way \citep{RS84}, the perturbation of the Oort cloud by a passing star \citep{HAE87}, or the existence of an undetected companion \citep{DHM84}.

If we assume that a kilometer-sized object could indeed strike an Earth-like planet, the energy $E_m$ deposited during the impact is
\begin{equation}
    E_m = \frac{1}{2} M_a \left(v_e^2 + v_\infty^2\right),
\end{equation}
where $M_a$ is the object's mass and $v_e = \sqrt{2 G M_p/R_p}$ is the escape velocity of the planet. Substituting the characteristic values, we find that $E_m$ falls within the range of $10^5-10^6$ Mt, which has been posited as the threshold at which environmental damage becomes global in scale \citep{TZM97}. Some of the predicted outcomes include the agglomeration of dust, soot and sulfates in the atmosphere (reducing the insolation), and the formation of nitrogen oxides due to shock impact and the consequent depletion of the ozone layer.

Hence, it is conceivable that interstellar impactors could serve as the ``pulse'' in the press-pulse model of mass extinctions \citep{AW08,GGH17}, with the ``press'' being supplied in tandem by geological processes such as volcanism. Moreover, the fossil diversity record (on Earth) suggests that high origination rates occur in the aftermath of high extinction rates \citep{Ben95,Al08}; ecological niches vacated by extinct species are quickly filled by new ones. Most notably, there is much evidence in support of the hypothesis that the adaptive radiation of Early Cenozoic mammals occurred in the wake of the K-Pg extinction that killed the dinosaurs \citep{Luo07,HUG17}; see, however, \citet{WE12}. 

As a result, it has been suggested that compact asteroid belts may be one of the prerequisites for the origin of complex life, and that such belts are likely to be uncommon \citep{ML13}. For such systems (sans compact asteroid belts) the impacts from interstellar objects may possibly serve as an alternative means of regulating macroevolutionary processes, albeit in a more sporadic fashion.

\subsection{The delivery of organic materials and life}
On Earth, there is fairly unambiguous evidence suggesting the existence of life at 3.8 Gyr ago \citep{KBS16}. It was previously thought that the Earth could not have hosted life during the Late Heavy Bombardment (LHB), but two factors suggest that this was not necessarily the case: (i) the existence of hyperthermophiles \citep{AM09}, and (ii) the LHB may not have been as localized as originally proposed \citep{BN17}. In addition, there exists some evidence, albeit subject to much ambiguity \citep{PTP18}, indicating the existence of life on Earth as early as $\sim 4.1$-$4.3$ Gyr ago \citep{BBHM,Dod17}. Thus, we will assume a timescale of $t_A \gtrsim 200$ Myr for the origin of life (abiogenesis) on Earth, but this choice is merely a fiducial value. The frequency of impacts with the Earth as a function of the impactor size is determined from (\ref{numbden}) and (\ref{CollRate}). We find that $\sim 400$ interstellar objects of radius $\sim 0.1$ km could have struck the Earth prior to abiogenesis, while the corresponding number of km-sized objects is $\sim 10$. Hence, life could have been transferred to the Earth by means of lithopanspermia \citep{Bur04,Wick10}.

Several studies have investigated the feasibility of interstellar panspermia \citep{Mel03,AS05}, and recent numerical simulations appear to suggest that lithopanspermia between members of the Solar birth cluster was feasible \citep{BM12}. Assessing the biological survival of alien microorganisms within interstellar rocks is not possible since we do not know their biological survival limits nor the travel time. However, as seen from Tables VIIIa and VIIIb of \citet{MC00}, interplanetary panspermia between Mars and Earth could deliver as many as $\sim 10^{12}$ microbes in meter-sized objects (with suitable shielding) for transit times of $\lesssim 1$ Myr. Hence, it seems plausible that much larger objects, such as the ones discussed above, could transfer alive microorganisms; in fact, \citet{WM04} proposed that even a few kilograms of microbe-bearing fragments may suffice to seed the target planetary systems with life.

Comets and meteorites played an important role in our Solar system by transporting organic molecules to Earth \citep{EC00,THCM06}; the delivery of these biomolecules (pseudo-panspermia) imposes less stringent requirements than panspermia \citep{Linga}. We can estimate the mass of amino acids delivered per year $\dot{M}_A$ via interstellar objects of $\sim 100$ m by using (\ref{CollRate}) and the fact that amino acids may comprise $\sim 1\%$ of the total mass \citep{CTB90}; thus, we arrive at
\begin{equation}\label{MaRa}
    \dot{M}_A \sim 2 \times 10^2\,\mathrm{kg/yr},
\end{equation}
and this value is smaller than the solar-bound cometary impact delivery rate of $\sim 10^3-10^6$ kg/yr for Earth, although it is comparable to the delivery rate for meteorites \citep{CS92}. 

The mass of amino acids delivered over the timescale required for abiogenesis ($t_A \sim 200$ Myr) is estimated as $M_A \sim \dot{M}_A t_A$, which yields $M_A \sim 4 \times 10^8$ kg after making use of (\ref{MaRa}). The number of potentially life-bearing objects transported by means of lithopanspermia between nearest neighbors in a star cluster is $\sim 3 \times 10^8\, \ell$, assuming that $\ell$ km of the crust is ejected during the bombardment of the host planetary surface \citep{BM12}. Using the fact that the mass of each object was $> 10$ kg and assuming that amino acids comprise $1\%$ of the mass (as before), we end up with $M_L > 3 \times 10^7\,\ell$ kg, where $M_L$ is the total mass of amino acids transferred via lithopanspermia. It can be seen that $M_A$ and $M_L$ are roughly comparable to each other, implying that lithopanspermia in the birth cluster is an important factor that should be taken into consideration. 

Here, it must be noted that the calculation of $M_A$ and $M_L$ for our planet implicitly assumed that the origin of life occurred during the Hadean era; in general, the mechanism of lithopanspermia between members of the birth cluster is typically possible only when $t_A < 500$ Myr is valid \citep{BM12}. In this context, an important point should be recognized concerning our prior analysis. The capture rates estimated previously should probably be viewed as lower bounds for earlier epochs where planetary systems co-existed in their common birth cluster. This is because the higher density of objects exchanged between nearest neighbors can result in an enhanced capture rate as seen from (\ref{CapRate}).

We note that interstellar objects can traverse through spatial regions in close proximity to O/B-type stars and receive high (but transient) doses of UV radiation leading to the rapid formation of biologically relevant molecules \citep{Roop}. Interstellar objects might also transport biomolecules not readily available via exogenous delivery within our solar system. For example, meteorites on Earth apparently lack the nucleobases cytosine and thymine, thus making the RNA/DNA world problematic from this particular standpoint \citep{PP16}. In addition to delivering intact organics, the synthesis of prebiotic compounds could also be facilitated due to the energy released from the impacts \citep{DW10}. Many organic molecules - including amino acids, peptides and nucleobases - have been synthesized via this process \citep{THCM06,MP13,FNS15}. 

Prebiotic synthesis of biomolecules due to interstellar objects (either via exogenous delivery or shock impacts) is expected to be lower than the corresponding values for native asteroids and comets by a few orders of magnitude. Nevertheless, we anticipate that this pathway can become important for planetary systems that lack sufficient cometary material.

\subsection{Seeding life through capture}
Hitherto, we have restricted ourselves to discussing panspermia (or pseudo-panspermia) wherein the interstellar object directly impacts the Earth, or a habitable planet. However, there is another channel by which interstellar objects can serve as a means of spreading life. 

We had concluded previously that it is possible to capture objects of up to $\sim 60$ km within our Solar system. On the other hand, in the $\alpha$-Cen system (and binaries in general), the maximum size of the captured object over the system's history was predicted to be nearly equal to the Earth. Hence, in such circumstances, it is possible that the captured ``planet'' may have already developed life and could thus spread it to other planets by means of \emph{interplanetary} panspermia. Thus, the interstellar object does not directly impact another planet, but rather serves as the ``carrier'' and life is transported due to collisions with impactors and spewing out ejecta. 

The minimum radius of the impactor $R_i$ required for producing ejecta that escape into space is estimated from the following formulae \citep{MC00,AI04}:
\begin{equation}
    R_e \approx \frac{3 P_s}{8g\rho_c} \left(\frac{\Delta v + v_e}{\Delta v - v_e}\right),
\end{equation}
\begin{equation}
    \frac{R_e}{R_i} \approx 2 \times 10^{-3}\,\left(\frac{v_i}{10\,\mathrm{km/s}}\right)^{-2} \left(\frac{\Delta v}{v_i}\right)^{-2/3},
\end{equation}
where $R_e$ is the radius of the ejecta, $P_s$ and $g$ are the atmospheric pressure and surface gravity of the planet, $\rho_c$ denotes the density of the crust, while $\Delta v$ and $v_i$ denote the post-impact and impact velocities respectively. If we hold all parameters fixed except for $p$ and $g$, it is seen that $R_i \propto P_s/g$. As per our previous discussion, the captured object is likely to be smaller than the Earth even for binary stellar systems like $\alpha$-Cen. From our Solar system, it is evident that most moons and planets smaller than the Earth possess much lower surface pressures. Hence, for such worlds, we are led to conclude that a smaller impactor would typically suffice to produce ejecta, and consequently enable interstellar lithopanspermia. A similar line of reasoning was invoked by \citet{Hout11} to posit that Ceres may have seeded the terrestrial planets of our Solar system via this process (termed ``glaciopanspermia'').

We also note that the captured planet- or moon-sized interstellar object could also develop life from a previously frozen and uninhabited state. In particular, this could happen if the new orbit falls within the habitable zone such that the ice envelope melts to yield a liquid ocean and an atmosphere. Of course, this route presupposes the existence of sufficient  volatiles and bioessential elements \citep{Mana18,LL18c}.

\section{The ``interstellar fishing net'' of the Solar system}\label{SSecFish}
One of the chief implications arising from Sec. \ref{SecCap} is that there may be $\sim 6 \times 10^3$ interstellar objects currently surviving in the Solar system. Hence, searching for these objects presents a more viable alternative to sending out interstellar probes in order to study the debris from exoplanetary systems. The situation can be likened to that of a fishing net for catching fishes, with interstellar asteroids or comets representing the ``fishes'' and the Jupiter-Sun binary system serving as the ``fishing net'' that captures these objects.  

The notion that one can look for interstellar objects within our Solar system has also been explored widely in the context of the Search for Extraterrestrial Intelligence (SETI), except that the ``objects'' are extraterrestrial artifacts \citep{Fre83,HK12,Dav12,Wri18} that are either defunct or operational. We note that these artifacts could have either arrived by chance or sent deliberately, perhaps as a means of energy-efficient communication \citep{RW04,Arn13}.

An immediate issue that arises is the necessity for a means of distinguishing between captured objects from interstellar space and those within our Solar system \citep{Gai17}. It is well-known that the oxygen isotope ratios ($^{17}$O/$^{16}$O and $^{18}$O/$^{16}$O) for carbonaceous 
chondrites in our Solar system yield a slope of $\sim 1$ in the oxygen three-isotope plot, although some deviations do exist \citep{Clay93,Clay03,YK08}. Hence, if the oxygen isotope ratios are markedly different from the values commonly observed in the Solar system, it may suggest that the object is interstellar in nature; more specifically, the ratio of $^{17}$O/$^{18}$O is distinctly lower for the Solar system compared to the Galactic value \citep{NG12}, and hence a higher value of this ratio may be suggestive of interstellar origin. Apart from utilizing oxygen isotopes as diagnostic tools, other isotope ratios like $^{12}$C/$^{13}$C and $^{14}$N/$^{15}$N \citep{KA09,Mum11} might help further distinguish between the two classes of objects.\footnote{For example, if the object had originated in, or passed through, the vicinity of Young Stellar Objects (YSOs), the $^{12}$C/$^{13}$C isotope ratio could be higher than the value in the interstellar medium \citep{SPY15}.}

In principle, it should be possible to analyze some of these isotopes by means of high-resolution spectroscopy in the optical, infrared and submillimeter ranges of water vapor in cometary tails \citep{JM09}. For instance, the HIFI instrument on the Herschel Space Observatory was used to measure the $^{18}$O/$^{16}$O isotope ratio of the comet C/2009 P1 in the Oort cloud \citep{BBS12}. Sending a probe to flyby the interstellar object or land on it and retrieve samples will be more challenging \citep{SL18}, but the scientific advantages are expected to be commensurate \citep{HP17}.\footnote{In our Solar system, missions such as Stardust, Rosetta and EPOXI have yielded a great deal of data through flybys, landings or sample returns.} One of the chief auxiliary advantages of the recent \emph{Breakthrough Starshot} project, which relies on the development of light-sail propulsion, may therefore stem from its ability to undertake rapid (on the order of hours to days) flyby missions of interstellar objects within our Solar system.\footnote{\url{https://breakthroughinitiatives.org/initiative/3}}

In the case of captured interstellar objects, we observe that the additional methods described below may serve as a means of distinguishing them from the ones formed in our Solar system.
\begin{itemize}
    \item The distribution of semi-major axes for captured objects (denoted by $a_c$) was computed through a semi-analytic procedure in Fig. 5 of \citet{Tor86}. A significant fraction of captured objects were found to have $a_c \lesssim 10$ AU. The rest of the objects are expected to be characterized by $a_c \gtrsim 100$ AU \citep{NM18}.
    \item Although not a diagnostic by itself, numerical simulations suggest that most captured interstellar objects should be characterized by relatively high inclinations ($i_c$) and eccentricities ($e_c$). From Fig. 2 of \citet{NM18}, it can be seen that the captured objects with $a_c \lesssim 10$ AU have $i_c \sim 145^\circ$-$170^\circ$ and $e_c \sim 0.2$-$0.35$, whereas the objects at $a_c \gtrsim 100$ AU have $i_c \sim 45^\circ$-$135^\circ$ and $e_c \sim 0.5$-$0.9$. The asteroid (514107) 2015 BZ$_{509}$ has parameters consistent with this range of values \citep{WCV17}, but it remains unclear as to whether it may be of interstellar origin.
    \item Apart from the isotope ratios discussed earlier, anomalously low CN-to-OH ratios and abundances of C$_2$ and C$_3$ have also been hypothesized as markers for interstellar objects \citep{LS07,Sch08}.
\end{itemize}

\section{Conclusion}
We have found that it would be easier to detect smaller interstellar objects at closer distances by means of reflected sunlight, while larger objects at greater distances are more detectable through their thermal radiation due to radiogenic heating.

The calibrated number density of interstellar objects, based on the detection of `Oumuamua, allowed us to estimate the capture rate of such bodies by means of three-body interactions for both our Solar system and stellar binaries such as the nearby $\alpha$-Centauri. We have found that a few thousand captured interstellar objects might be found within the Solar system at any time. The largest of these would be an object with radius tens of km. For the $\alpha$-Centauri A\&B system, we have found that even Earth-sized objects could have been captured through this process and that Moon-sized objects may currently exist in circumstellar orbits. Our results depend to the power-law index for the size distribution of interstellar objects, which is currently unknown.

Our findings have potentially important implications for the origin and evolution of life on Earth. If a km-sized interstellar object were to strike the Earth, we suggested that it would result in pronounced local changes, although the global effects may be transient. Habitable planets could have been seeded by means of panspermia through two different channels: (i) direct impact of interstellar objects, and (ii) temporary capture of the interstellar object followed by interplanetary panspermia. There are multiple uncertainties involved in all panspermia models, as the probability of alien microbes surviving ejection, transit and reentry remains poorly constrained despite recent advancements.

The Solar system acts as a fishing net, enabling us to search for traces of extraterrestrial life locally (due to the presence of captured interstellar objects) as opposed to sending interstellar probes. The same approach is also applicable to the search for extraterrestrial artifacts within our Solar system. Possible methods of distinguishing between extrasolar and solar objects include the analysis of their orbital trajectories and oxygen isotope ratios, the latter of which can be ostensibly inferred from high-resolution spectroscopy of water vapor in cometary tails. A few months after the submission of our paper, an interstellar origin was suggested for the asteroid (514107) 2015 BZ$_{509}$ based on its documented retrograde motion in Jupiter's co-orbital region. This report \citep{NM18} highlights the timeliness and importance of our study.

\acknowledgments
We thank David Jewitt for valuable comments concerning the paper. This work was supported in part by grants from the Breakthrough Prize Foundation for the Starshot Initiative and Harvard University's Faculty of Arts and Sciences, and by the Institute for Theory and Computation (ITC) at Harvard University.

%\bibliographystyle{aasjournal}
%\bibliography{Oum}

\end{document}